\documentclass[prd,twocolumn,showpacs]{revtex4}
\usepackage{amssymb}
\usepackage{amsmath}
\usepackage{graphicx}
\begin{document}

\title{Analysis to the entangled states from an extended Chaplygin gas model%
}

\author{Xin-He Meng$^{1,2}$}
\altaffiliation{Corresponding author}
\email{xhm@nankai.edu.cn}
\author{Ming-Guang Hu$^{2}$}
\email{huphys@hotmail.com}
\author{Jie Ren$^{2}$}
\affiliation{$^{1}$CCAST (World Lab), P.O.Box 8730, Beijing 100080, China\\
$^{2}$Department of physics, Nankai University, Tianjin 300071, China}

\begin{abstract}

With considerations of the recently released WMAP year three and supernova
legacy survey (SNLS) data set analysis that favors models similar to the $%
\Lambda CDM$ model by possibly mild fluctuations around the vacuum energy or
the cosmological constant, we extend the original Chaplygin Gas model (ECG)
via modifying the Chaplygin Gas equation of state by two parameters to
describe an entangled mixture state from an available matter and the rest
component (which can take the cosmological constant or dark energy as in the
current cosmic stage, or `curvature-like' term, or radiation component in
the early epoch, as various phases) coexistence. At low redshifts, the
connection of the ECG model and the Born-infeld field is set up. As
paradigms, we use the data coming from the recently released SNLS for the
first year and also the famous 157 type Ia supernova (Ia SNe) gold dataset
to constrain the model parameters. The restricted results demonstrate
clearly how large the entangled degree or the ratio between the energy
density parameters of the two entangled phases being. The fact that the ECG
models are consistent with the observations of Ia SNe is obtained through
the redshift-luminosity distance diagram, hence the ECG can be regarded
possible candidates for mimicking the current speed-up expansion of our
universe.
\end{abstract}

\pacs{98.80.Cq}
\maketitle


\section{Introduction}

\label{sec-intro} 
The recently released high redshift SNe (SNLS-Supernova Legacy
Survey project) data set has been analyzed to show it, in better
agreement with WMAP year three CMB observations, favors the
$\Lambda CDM$ model but with its equation of state mildly around
the very -1 as for a preferred cosmological constant \cite{pa}. On
the other side that the serious fine-tuning problem for the
cosmological constant problem and the dark side physics of
Universe that have been puzzling us across the century \cite{sc},
especially the recent years discovery that our universe is
undergoing an accelerating phase, maybe due to a mysterious
component as coined Dark Energy, forces us to model the perplexing
situation more realistically, that is we try to reconcile the
simple cosmological constant by a dynamically slowing varying
composite with limit case back to the economic cosmological
constant. We know the observational fact that our current universe
evolution is controlled by a mixture of dark energy (we know
less), matter (mainly the dark matter, knowing equally poor, even
not less) and radiation (knowing better except for the neutrino's
absolute mass) or curvature contribution (uncertain existence) or
other possible form of fluid (not certain yet). Observations of
type Ia supernova(SNe Ia) directly suggest that the expansion of
the universe is accelerating with possibly powered by the dark
energy, and the measurement of the cosmic microwave background
(CMB)
\cite{DNS} as well as the galaxy power spectrum for large scale structure%
\cite{MT} indicate that in a spatially flat isotropic universe, about
two-thirds of the critical energy density seems to be stored in a the dark
energy component with enough negative pressure responsible for the currently
cosmic accelerating expansion\cite{AGR}. It is clear from observations that
most of the matter in the Universe is in a dark (non-baryonic) form (see,
for instance, \cite{pd}). To 
understand the cosmic dark component physics is certainly helpful for us to
better model our universe and a great challenge for temporary physicists.

The simplest candidate for the dark energy is a cosmological constant $%
\Lambda $, which has a specially simple pressure expression
$p_{\Lambda }=-\rho _{\Lambda }$. However, the $\Lambda $-term
requires that the vacuum energy density be fine tuned to have the
observed very tiny value, the famous \textquotedblleft\ old"
cosmological constant problem. To alleviate this, many other
different forms of dynamically changing dark energy models have
been proposed instead of the only cosmological constant
incorporated model, such as modified gravity models \cite{xh}.
Usually, the equation of state (EOS) for describing dark energy
can be assumedly factorized into the form of $p_{DE}=w\rho _{DE}$,
where $w$ may depend on cosmological redshift $z$ \cite{DP}, or
scale factor $a(t)$ with a more complicated parametrization. The
case for $w=-1$ corresponding to the cosmological constant, was
thought as a border-case as named the phantom divide in \cite{WH}.

Recent years a new kind of model called Chaplygin Gas model (CGM)
from statistic physics with its special EOS \cite{Kamen(2001)} as
\begin{equation*}
p=-\frac{A}{\rho }
\end{equation*}%
leads to a density evolution of the form
\begin{equation*}
\rho =\sqrt{A+\frac{B}{a^{6}}}.
\end{equation*}%
It interpolates between matter at relatively early epoch and dark
energy/cosmological constant at late stage. However, it would suffer
problems when considering structure formation \cite{Sandvik(2004)} and
cosmological perturbation power spectrum.

Later, a modified Chaplygin Gas Model, by generalized Chaplygin
gas model
(GCGM) with an EOS as%
\begin{equation*}
p=-\frac{A}{\rho ^{\alpha }}
\end{equation*}%
has been discussed largely in the Ref. \cite{GCG} with the
motivation to overcome the original model shortcomings. It
describes a broad class of universe models including CGM, for
choosing different range for the uncertain parameter, $\alpha $,
with the energy density expressed formally as:
\begin{equation}
\rho =\left[ A+\frac{B}{a^{3(1+\alpha )}}\right] ^{1/(1+\alpha )}=\left[
\rho _{\Lambda }^{\alpha +1}+\rho _{m}^{\alpha +1}\right] ^{1/(1+\alpha )},
\label{eq-1-GCGdens}
\end{equation}%
in which both $A$ and $B$ are parameters and, $0\leq \alpha \leq
1$. When one takes $\alpha =1$, it returns to the original CGM. It
is remarkable that the model interpolates like the CGM between a
de Sitter universe and a dust-dominated one via an intermediate
phase which is a mixture of a cosmological constant and a perfect
fluid with a \textquotedblleft\ soft" matter equation of state,
$p=\alpha \rho $ \cite{Bento(1999)}. This mixed state is not the
same as conventional ones, for its non-linear mixture property in
the right hand side of the Friedmann equation
\begin{equation}
\rho \neq \sum \rho _{i}=\rho _{m}+\rho _{DE}+\rho _{k}+\cdots ,
\label{eq-H2}
\end{equation}%
with these suffixes indicating possible contributions from matter, dark
energy, curvature, ...


This is just one choice to extend the CGM. For the parameter $A$ in CGM, it
can be generalized, too \cite{mh}. A natural and simple consideration for $A$
is a power-law relation with the redshift, $z$ (in the standard cosmology $%
z=1/a-1$), e.g. $A(z)=A_{0}(1+z)^{m}$. Based on GCGM, such extension would
give a model with a bigger parameter space-$\{m,\alpha \}$ (The case $m=0$
leads back to GCGM). Directly from Eq. (\ref{eq-1-GCGdens}) may a potential
merit of such extension that the density evolution has got a little more
general form
\begin{equation}
\rho =\left[ A_{0}(1+z)^{m}+\frac{B}{a^{3(1+\alpha )}}\right] ^{1/(1+\alpha
)},  \label{eq-1-ECGdens}
\end{equation}%
a non-linear combination between two different fluids density be
palpably recognized. At the low redshift cases the first term in
the square bracket can be naturally tackled as a small
perturbation to the constant $A_{0}$ and the above ansatz also
implies that such energy density can mimick a mixed state between
a dust-dominated phase ($\rho \propto a^{-3}$) and an optional one
like cosmological constant, the radiation-like ($\rho \propto
a^{-4}$) or curvature-like phase ($\rho \propto a^{-2}$)
determined by proper choosing the value of $m$. We may as well use
the notion of entanglement which mostly appears in quantum
mechanics to denote the above mixed phenomenon, since energy
density that depicts the stage at what the universe is has had its
function like eigenstate in quantum mechanics. Especially, the
modern astrophysical observations including Type Ia Supernovae,
Cosmic Microwave Background, Large Scale Structure, etc. endue us
with the chance to test whether this model is a plausible
candidate for explaining both the earlier matter dominated phase
and recent cosmic speed-up expansion of the universe. This
motivates us to explore such Extended Chaplygin gas model
abbreviated as ECGM during our following discussions.

The paper is arranged as follows: In Sec. \ref{sec:modified chaplygin gas}
the general forms of the ECGM are introduced and in Subsec. \ref{subsec:m=3n}
and \ref{subsec:m<3n} two types of the ECG models are investigated
respectively; the parameter constraints and some discussions are arranged in
the Sec. \ref{sec: investigation}. At the end, we present our conclusions
with some discussions.

\section{Extended Chaplygin gas model}

\label{sec:modified chaplygin gas} A homogeneous, isotropic and flat
Robertson-Walker metric can be written as
\begin{equation}
ds^{2}=-dt^{2}+a^{2}(t)(dx^{2}+dy^{2}+dz^{2}),
\end{equation}%
where $a(t)$ is the expansion scale factor. Under the framework of
Friedman-Robertson-Walker cosmology, the global dynamic evolution of the
universe is manipulated by the Friedmann equations
\begin{eqnarray}
\left( \frac{\dot{a}}{a}\right) ^{2} &=&\frac{8\pi G}{3}\rho ,
\label{2:Fri1} \\
\frac{\ddot{a}}{a}+\left( \frac{\dot{a}}{a}\right) ^{2} &=&-4\pi G(\rho +p),
\label{2:Fri2}
\end{eqnarray}%
and the energy-conservation equation
\begin{equation*}
\dot{\rho}=-3H(\rho +p),
\end{equation*}%
or its equivalent form
\begin{equation}
d(\rho a^{3})=-pd(a^{3}).  \label{2:energy conservation}
\end{equation}%
%
The unit convention $8\pi G/3=c=1$ is used in this paper, the dot denotes
the derivative to the time, and the symbol $H$ represents the Hubble
parameter while $z$ the redshift throughout. Moreover, the present scale
factor $a_{0}$ is also assumed to be unit for brevity.

At the right side of the Eq. (\ref{2:Fri1}) there are often two conventional
treatments:

\begin{itemize}
\item the total energy density is usually a linear addition of various
fluids density, like%
\begin{equation*}
H^{2}=\rho _{m}+\rho _{DE}+\rho _{k}+\cdots ,
\end{equation*}%
with meanings of these suffixes as matter, Dark energy, curvature, ...

\item For the mysterious dark energy, one common way treating it
is to fix the matter component but proposing different models or
analysis for the dark energy and other possible parts.
\end{itemize}

In this paper, we will violate these above two conventions to
investigate the dark (matter and energy) fluid universe below.

From the viewpoint of the quantum field theory, the vacuum energy (or
cosmological constant $\Lambda $) is the eigenvalue of the ground state of a
quantum field. It is admissible to set vacuum energy a non-zero constant in
the ECG model. Equivalently there are two energy fluids, the dynamical part,
which can be regarded as a fluctuation and the cosmological constant part
(it violates the second convention). Consequently, it deduces the following
relations:
\begin{equation*}
\rho _{ECG}=\rho _{\mathrm{v}}+\rho _{dyn},\qquad p_{ECG}=p_{\mathrm{v}%
}+p_{dyn},
\end{equation*}%
in which the suffix, $\mathrm{v}$, represents vacuum and $dyn$,
dynamical contributions respectively. The EOS of the two kinds of
fluids are
\begin{equation}
p_{\mathrm{v}}=-\rho _{\mathrm{v}},\qquad p_{dyn}=-\frac{A_{0}(1+z)^{m}}{%
\rho _{dyn}^{\alpha }},  \label{2:EOS}
\end{equation}%
where parameters $p_{dyn}$ and $\rho _{dyn}$, respectively, denote the
pressure and the energy density of the dynamically evolving component of the
ECG media. [For the sake of simplicity, we substitute $p_{dyn}$ and $\rho
_{dyn}$ simply with $p$ and $\rho $ in the succeeding sections.]

In this model, we adopt the total effective EOS parameter $w_{T,eff}$ as
defined in \cite{Linder(2004)}
\begin{equation}
w_{T,eff}(z)\equiv -1+\frac{1}{3}\frac{d\ln (H^{2}/H_{0}^{2})}{d\ln (1+z)}.
\label{2:statePa}
\end{equation}%
The sound speed $c_{s}$ defined by $c_{s}^{2}=p_{ECG}^{\prime }/\rho
_{ECG}^{\prime }$ here becomes
\begin{equation}
c_{s}^{2}(z)=-\alpha \frac{p}{\rho }+\frac{m}{1+z}\frac{p}{\rho ^{\prime }},
\label{eq:sound}
\end{equation}%
for $p_{\mathrm{v}}$ and $\rho _{\mathrm{v}}$ are constants. Primes denote
the derivative about the redshift $z$ with $m\neq 0$. As $m=0$, it returns
to the GCG case. The first term in the right hand is just the sound speed
for GCG model, while the second in the lower redshift can be treated as a
small perturbation which we mentioned above. We will check below whether
this sound speed would exceed the speed of light (see Ref. \cite{DeDeo(2003)}
etc.).

The relationship between density $\rho (a)$ and scale factor $a$ is derived
by means of Eqs. (\ref{2:energy conservation}) and (\ref{2:EOS}). That is,
\begin{equation*}
\int d\left[ a^{3(\alpha +1)}\rho ^{(\alpha +1)}\right] =3(\alpha
+1)A_{0}\int a^{3\alpha +2}(1+z)^{-m}da,
\end{equation*}%
or a more applicable form
\begin{equation}
\int d\left[ a^{3(\alpha +1)}\rho ^{(\alpha +1)}\right] =3(\alpha
+1)A_{0}\int a^{3\alpha -m+2}da.  \label{2:pp}
\end{equation}%
In terms of the right hand side of the Eq. (\ref{2:pp}), we can
mathematically divide the model into two classes according to whether the
equation $m=3(\alpha +1)$ can be satisfied or not.

\subsection{The $m=3(\protect\alpha+1)$ case}

\label{subsec:m=3n} 
In this subsection, the relation $m=3(\alpha +1)$ holds and thus the Eq. (%
\ref{2:pp}) gives the following result:
\begin{subequations}
\begin{equation}
\rho =\left[ \frac{A\ln a}{a^{3(\alpha +1)}}+\frac{B}{a^{3(\alpha +1)}}%
\right] ^{1/(\alpha +1)},  \label{2.1:rho-a}
\end{equation}%
and its equivalent form:
\begin{equation}
\rho =\left[ B-A\ln (1+z)\right] ^{1/(\alpha +1)}(1+z)^{3},
\label{2.1:rho-z}
\end{equation}%
where the $B$ is an integration constant and $A$ is expressed here
as
\end{subequations}
\begin{equation*}
A=3(\alpha +1)A_{0}.
\end{equation*}%
%
By introducing Eq. (\ref{2.1:rho-a}) into Friedmann equation (\ref{2:Fri1}),
we obtain
\begin{equation*}
\int \frac{a^{1/2}da}{(A\ln a+B)^{1/2(\alpha +1)}}=\int dt,
\end{equation*}%
which has not easily got an analytical solution. By virtue of the numerical
integration, the evolution scale factor complies the expanding scenario but
with a faint speed-up, which resembles that of the Einstein-de Sitter
universe.

\begin{figure}[tbp]
\includegraphics{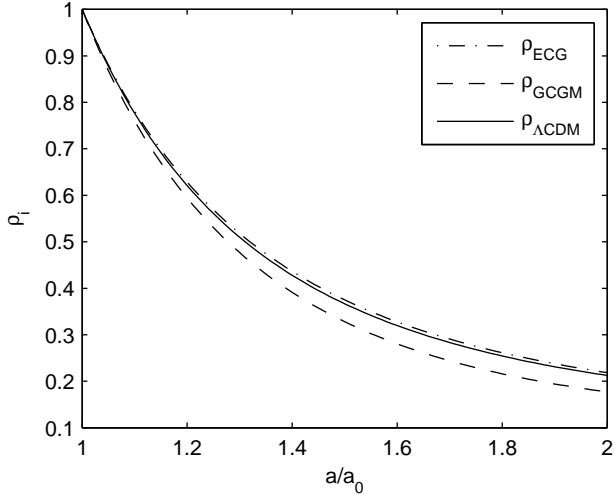}\newline
\caption{The energy densities of three different models are presented at
together at the low redshifts cases (We have assumed the energy density $%
\protect\rho_0$ at present days to be unit and the vacuum density to be $%
\protect\rho_\Lambda=0.1\protect\rho_0$).}
\label{fig:ECG1}
\end{figure}

That at the cases of lower redshifts the ECG model is actually the $\Lambda $%
CDM model plus a correcting-term can be understood from either the Eqs. [(%
\ref{2.1:rho-a}), (\ref{2.1:rho-z})] or the succeeding comparisons. We
display for convenience the energy density of the three models of $\Lambda $%
CDM, GCGM, and ECGM at the Tab. \ref{tab:table1}. Moreover, a trivial depict
of the energy density has been put on the Fig. \ref{fig:ECG1}. Clearly we
can see the $\Lambda $CDM can be described effectively by the ECGM with the $%
m=3(\alpha +1)$ at a sufficient degree of accuracy. In particular, we can
discuss the effective state parameter of this approximate model to the $%
\Lambda $CDM ($w=-1$) if one introduces the Eq. (\ref{2.1:rho-z}) back to
the Eq. (\ref{2:statePa}). As a result, the effective state parameter is
\begin{equation}
w_{T,eff}(z)=-1+\left[ 1-\frac{A_{0}}{B-A\ln (1+z)}\right] \frac{\rho }{%
H_{0}^{2}},  \label{2.1:statePa}
\end{equation}%
At low redshifts, one nontrivial point is that $A_{0}=B$ leading to $%
w_{T,eff}=-1$. Thus the second term at the right side can be viewed as a
small perturbation around $-1$.

\begin{figure}[tbp]
\includegraphics[width=8cm]{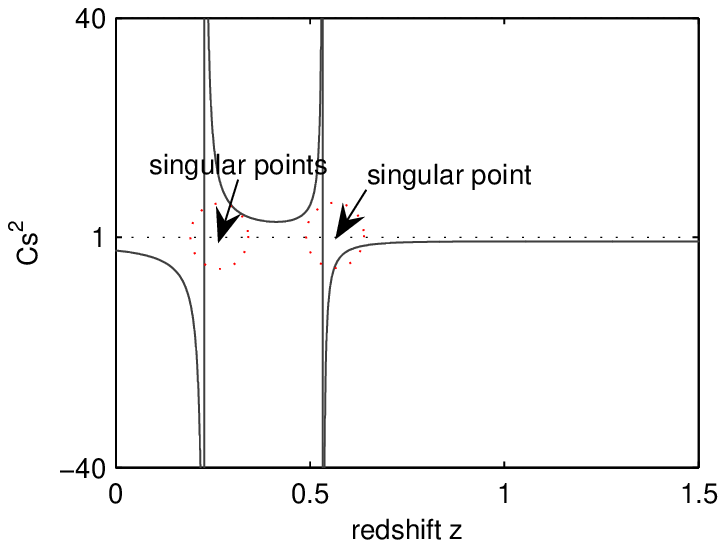} \includegraphics[width=8cm]{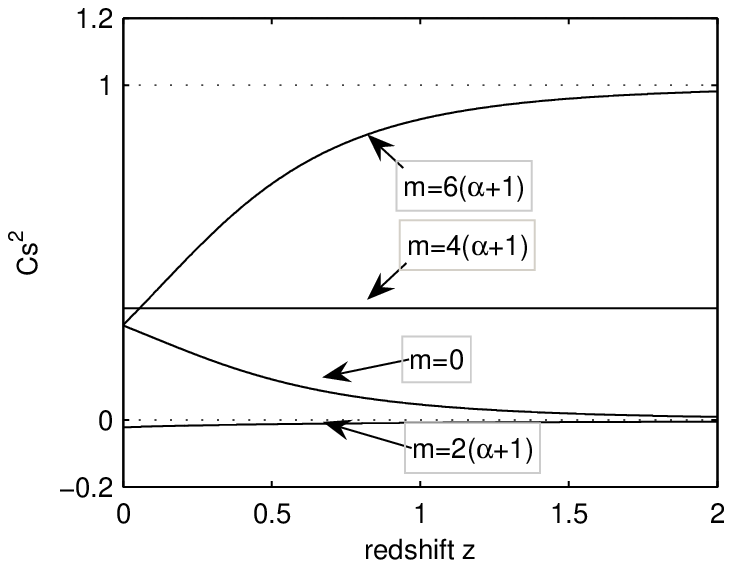}
\caption{The upper panel is the sound speed diagram of $m=3(\protect\alpha%
+1) $ and the lower one is that of $m\neq3(\protect\alpha+1)$. In the case
of $m=3(\protect\alpha+1)$, parameters take $A=0.7$, $B=0.3$ and $\protect%
\alpha=2/5$. In the case of $m\neq3(\protect\alpha+1)$, parameters take
values at Tab. \protect\ref{tab2}. }
\label{fig:Cs2}
\end{figure}

The sound speed Eq. (\ref{eq:sound}) becomes
\begin{equation*}
c_{s}^{2}(z)=-\alpha \frac{p}{\rho }\left[ 1+\frac{(\alpha +1)\left( B-A\ln
(1+z)\right) }{\alpha A_{0}-\alpha \left( B-A\ln (1+z)\right) }\right] ,
\end{equation*}%
Its property has been demonstrated on Fig. \ref{fig:Cs2}. Except
for the two singular points, the sound speed has got its values
less than $1$ (i.e, light speed) satisfying the causality.

Moreover, the similarity between such class of ECG models and the $\Lambda $%
CDM model may indicate that the Tachyon scalar field ( see reference \cite%
{Padmanabhan(2002)}) that can depict properties of the $\Lambda
$CDM is a possible source to characterize this class of ECG models
as well.
\begin{table*}[tbp]
\caption{This is a table which is filled with the energy density of the
three different models for the purpose of comparison. }
\label{tab:table1}%
\begin{ruledtabular}
\begin{tabular}{lcr}
$\Lambda$CDM&GCGM&ECGM\\
\hline $\rho=A(1+z)^3+\rho_{\Lambda},$&$
\rho=\left[\rho_\Lambda^{(1+\alpha)}+B(1+z)^{3(1+\alpha)}\right]^{1/(1+\alpha)}
,$&$ \rho=[B-A\ln (1+z)]^{1/(\alpha+1)}(1+z)^{3}+\rho_\Lambda.$
\end{tabular}
\end{ruledtabular}
\end{table*}

\subsection{The $m\neq3(\protect\alpha+1)$ case}

\label{subsec:m<3n} Now we elucidate the rest class of ECG models which
persist the inequation, $m\neq 3(\alpha +1)$. Thus, the Eq. (\ref{2:pp})
gives the following result
\begin{subequations}
\begin{equation}
\rho =\left[ \frac{A}{a^{m}}+\frac{B}{a^{3(\alpha +1)}}\right] ^{1/(\alpha
+1)},  \label{2.2:rho-a}
\end{equation}%
or its equivalent form
\begin{equation}
\rho =\left[ A(1+z)^{m}+B(1+z)^{3(\alpha +1)}\right] ^{1/(\alpha +1)},
\label{2.2:rho-z}
\end{equation}%
where $A$ is used to denote
\end{subequations}
\begin{equation*}
A=\frac{3(\alpha +1)A_{0}}{3(\alpha +1)-m}.
\end{equation*}%
One worth noticing point lies in the reality that when $m$=0, it returns to
the GCG model while to $\Lambda $CDM as $m=3(\alpha +1)$ or $A=0$. The
clarity that the energy density evolution relying on two nontrivial parts,
the dust component (the second term in the square bracket) and the uncertain
part (the first term in the square bracket) is presented by the density
formula. It also implements the density form Eq. (\ref{eq-1-ECGdens}) that
we have expected in the Sec. \ref{sec-intro}. At this case, proper choosing $%
m$ values can ascertain which phase is entangled with the non-relativistic
matter governed by the Eq. [ (\ref{2.2:rho-a}) or (\ref{2.2:rho-z})]. In the
Sec. \ref{sec: investigation}, we will use the low redshifts data from type
Ia supernova observations along with other kind of data to restrict such
entangled states. However, before that we first of all formally present some
connections, especially with the complex scalar field.

The EOS in this case from the Eq. (\ref{2:EOS}) becomes
\begin{equation}  \label{2.2:EOS}
w_{eff}=-1+\left[\frac{Am}{3(\alpha+1)}(1+z)^m+B(1+z)^{3(\alpha+1)}\right]%
/\rho^\alpha H_0^2.
\end{equation}

It is easy by analogus to quantum mechanics to define the entangled degree
as
\begin{equation}
P=\frac{2|A||B|}{A^{2}+B^{2}}.  \label{eq:degre}
\end{equation}%
For clarity, $P=0$ corresponds to no entanglement, while $P=1$ to the
maximal entanglement case.

The sound speed from Eq. (\ref{eq:sound}) turns out to be
\begin{equation*}
c_{s}^{2}=\frac{\alpha (1-Y)}{1+(1+z)^{3(\alpha +1)-m}/\eta }-\frac{m(1-Y)}{%
3Y+3(1+z)^{3(\alpha +1)-m}/\eta },
\end{equation*}%
where $Y$ takes
\begin{equation*}
Y=\frac{m}{3(\alpha +1)}.
\end{equation*}%
Inserting the fitted result at Tab. \ref{tab2}, the sound speed diagram has
been shown as in Fig. \ref{fig:Cs2}. For $m=2(\alpha +1)$ case, the sound
speed is negative which means (from the point of view for evolution of a
small perturbation as a wave $\ddot{\delta}-c_{s}^{2}\nabla ^{2}\delta
\simeq 0$) that collapsing regions and voids get amplified. For other cases,
the speeds are all positive and less than $1$.

\section{investigations about entanglement}

\label{sec: investigation} In this section, we have constrained the
different ECG models characterized with corresponding values of the
parameter $m$ by using the recently released SNLS SNe and nearby dataset
\cite{Astier(2005)} and the famous 157 SN Ia gold dataset \cite{Riess(2004)}%
). 

\subsection{data fitting}

The dimensionless Hubble parameter for the ECG model from the Eq. (\ref%
{2.2:rho-z}) reads
\begin{eqnarray}
E^2(z)&=&H^2/H_0^2=\rho_{ECG}/H_0^2  \notag \\
&=&\left[\left(\frac{\Omega_X}{\Omega_M}\right)^{\alpha+1}(1+z)^m+(1+z)^{3(%
\alpha+1)}\right]^{\frac{1}{\alpha+1}}\Omega_M  \notag \\
&&+\Omega_V,  \label{3-E}
\end{eqnarray}
where the symbols $\Omega_X=A^{1/(\alpha+1)}/H_0^2$, $\Omega_M=B^{1/(%
\alpha+1)}/H_0^2$, and $\Omega_V=\rho_\mathrm{v}/H_0^2$ represent
the present energy density parameters as defined by,
$\Omega_i=\rho_i/H_0^2$ ($i$ takes all possible components in
question), and $\Omega_i$ reflects the relative strength of the
component-$i$ at present.  These parameters together satisfy the
normalization condition as favored by observations:
\begin{equation*}
\left[\left(\frac{\Omega_X}{\Omega_M}\right)^{\alpha+1}+1\right]^{\frac{1}{%
\alpha+1}}\Omega_M+\Omega_V=1.
\end{equation*}

 From the analysis in the Subsec. \ref{subsec:m<3n}, we are aware
of that as
the parameter $m$ takes values--$0$, $2(\alpha +1)$, $4(\alpha +1)$, and $%
6(\alpha +1)$, the undetermined X-component of the energy density (\ref%
{2.2:rho-a}) describes, respectively, cosmological constant,
curvature term, radiation contribution, and stiff matter.
Hereafter, symbol $\eta $ is used to denote $\eta=\Omega
_{X}/\Omega _{M}$, and its significance can be shown in the
effective state parameter and entangled degree expression.

The most consistent values of $\Omega _{X}$, $\Omega _{M}$ and $\Omega _{V}$
are usually obtained through the cosmological fitting which is actually
performed in this work by minimizing
\begin{equation}
\chi ^{2}=\sum_{i}\frac{[\mu _{obs}(z_{i})-5\log _{10}(d_{L}(z_{i})/10%
\mathrm{pc})]^{2}}{\sigma _{i}^{2}+\sigma _{int}^{2}},  \label{3-chi-SN}
\end{equation}%
where $\sigma _{i}$ is the uncertainty in the individual distance moduli and
the $\sigma _{int}$ is the dispersion in supernova redshift (transformed to
units of distance moduli) due to peculiar velocities. In the flat universe,
the luminosity distance $d_{L}$ is defined by
\begin{equation*}
d_{L}(z)=\frac{1+z}{H_{0}}\int_{0}^{z}\frac{dz^{\prime }}{E(z^{\prime })},
\end{equation*}%
We set the Hubble constant
$H_{0}=75\mathrm{kms}^{-1}\mathrm{Mpc}^{-1}$ in accordance to that
appeared for SNLS SNe and nearby data in Ref. \cite
{Astier(2005)}.

It does not lose the generality for us to take in order $\alpha =1/2,2/3,4/5$
during the subsequent chi-square fit, which is compatible with the result in
the Ref. \cite{Bento(2003)}, that is, $0.2\lesssim \alpha \lesssim 0.6$.

\begin{figure}[tbp]
\includegraphics[width=4.2cm]{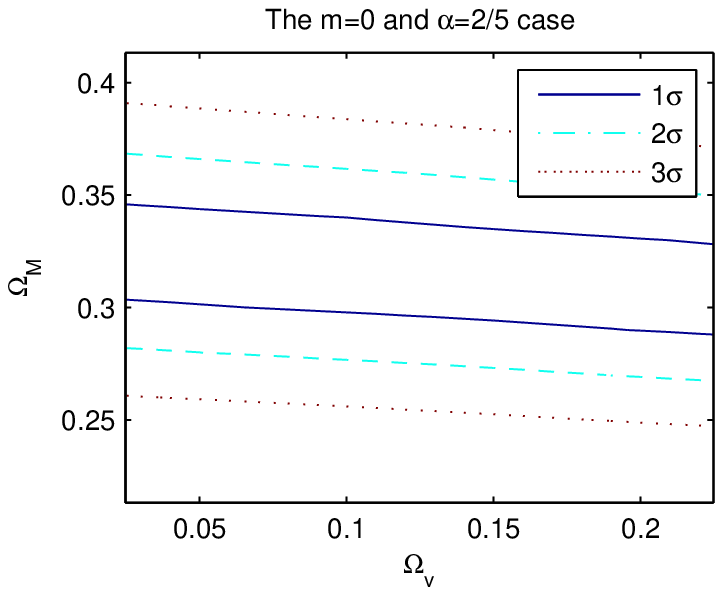} %
\includegraphics[width=4.2cm]{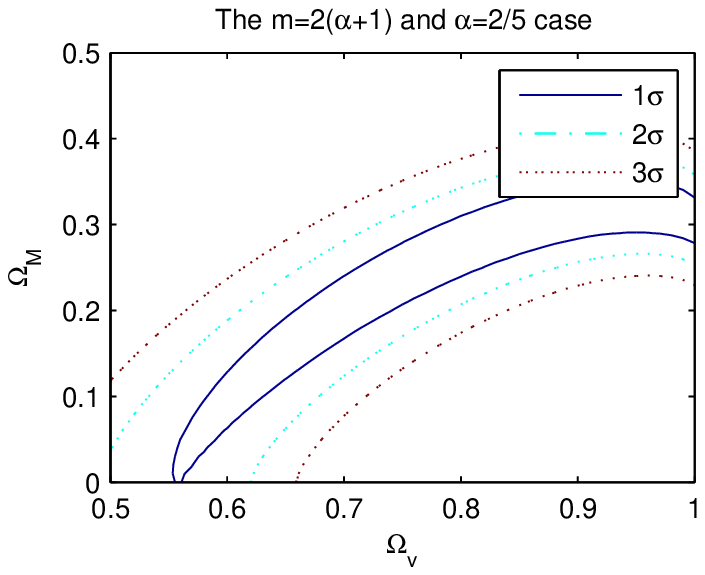} %
\includegraphics[width=4.2cm]{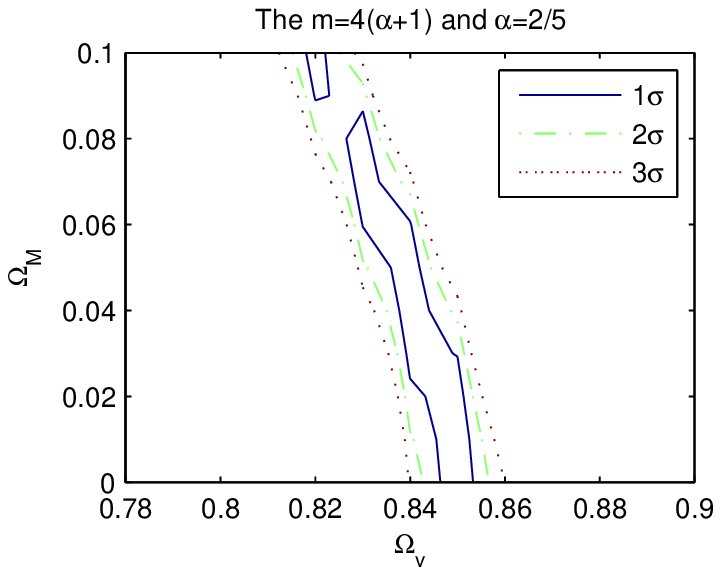} %
\includegraphics[width=4.2cm]{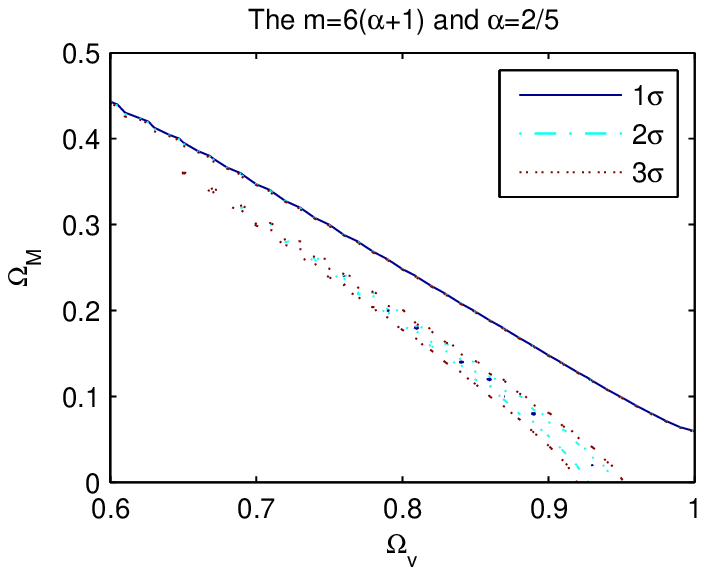}
\caption{The contours on the $\Omega_V-\Omega_M$ panel are in the case of $%
\protect\alpha=2/5$ for SNLS SNe and nearby data.}
\label{fig-SNLS1}
\end{figure}

\begin{table}[tbp]
\caption{The fitting results is in the case of $\protect\alpha=2/5$ for SNLS
SNe and nearby data. }
\label{tab2}%
\begin{ruledtabular}
\begin{tabular}{c|ccccc}
$m$&$\Omega_V$&$\Omega_M$&$\Omega_X$&$\chi^2$&$\eta$\\
\hline
$0$&$0.08$&$0.32$&$0.76$&$150.44$&$2.39$\\
$2(\alpha+1)$&$0.75$&$0.24$&$0.032$&$150.78$&$0.13$\\
$4(\alpha+1)$&$0.85$&$0$&$0.15$&$150.26$&$\infty$\\
$6(\alpha+1)$&$0.81$&$0.18$&$0.03$&$150.12$&$0.16$
\end{tabular}
\end{ruledtabular}
\end{table}

\begin{figure}[tbp]
\includegraphics[width=4.2cm]{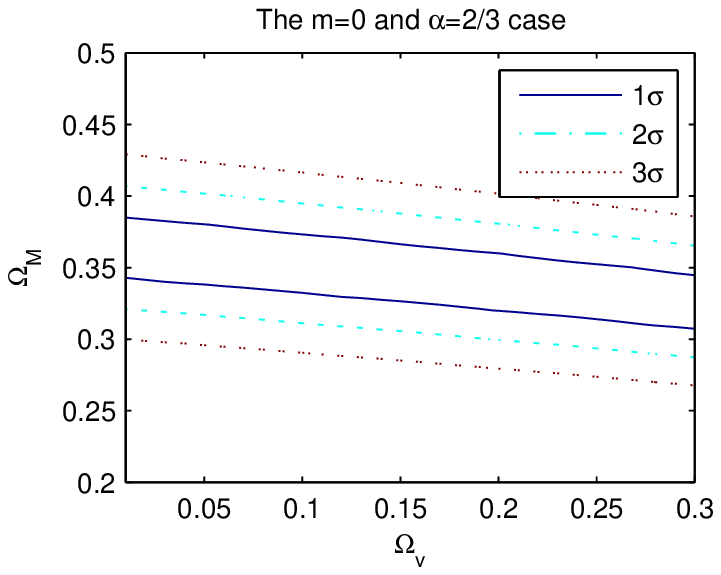} %
\includegraphics[width=4.2cm]{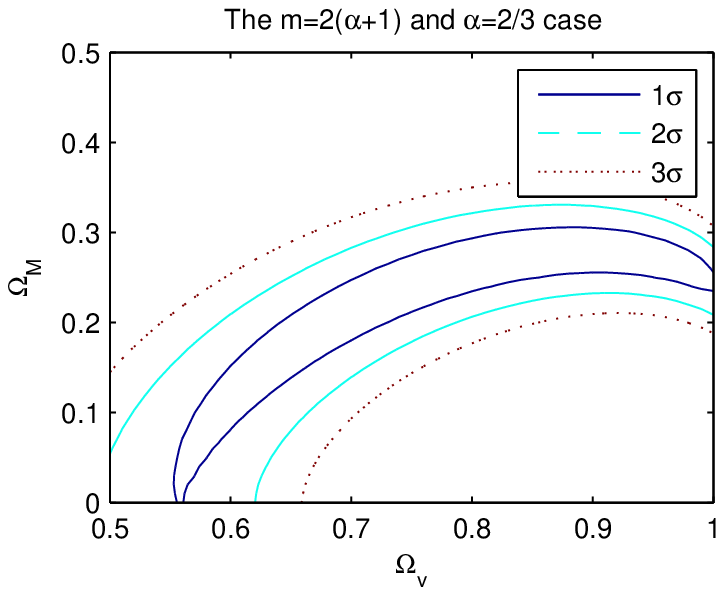} %
\includegraphics[width=4.2cm]{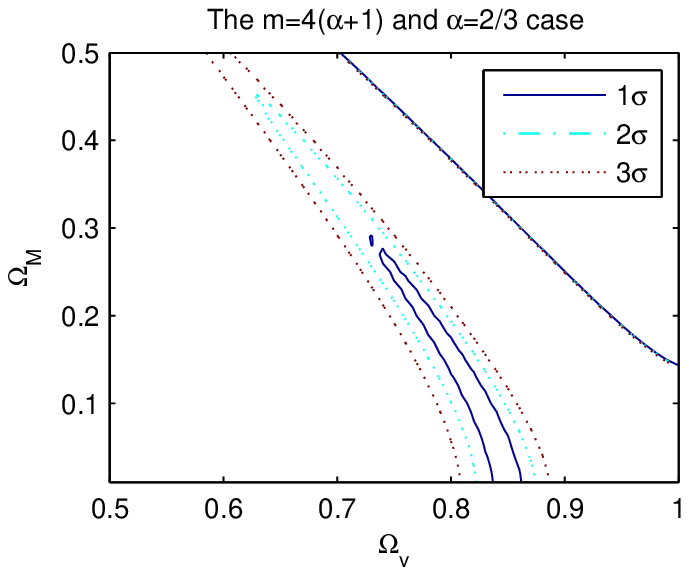} %
\includegraphics[width=4.2cm]{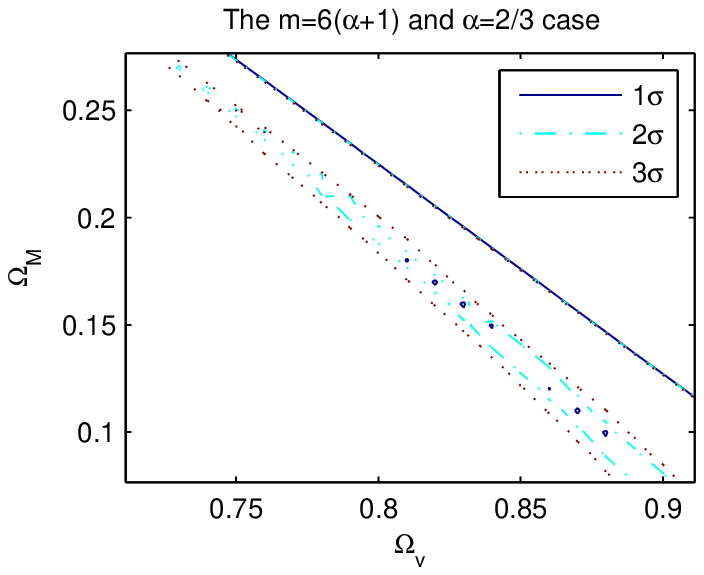}
\caption{The contours on the $\Omega_V-\Omega_M$ panel are in the case of $%
\protect\alpha=2/3$ for SNLS SNe and nearby data.}
\label{fig-SNLS2}
\end{figure}

\begin{table}[tbp]
\caption{The fitting results is in the case of $\protect\alpha=2/3$ for SNLS
SNe and nearby data. }
\label{tab3}%
\begin{ruledtabular}
\begin{tabular}{c|ccccc}
$m$&$\Omega_V$&$\Omega_M$&$\Omega_X$&$\chi^2$&$\eta$\\
\hline
$0$&$0.04$&$0.36$&$0.84$&$150.38$&$2.34$\\
$2(\alpha+1)$&$0.75$&$0.24$&$0.049$&$150.78$&$0.203$\\
$4(\alpha+1)$&$0.85$&$0$&$0.15$&$150.26$&$\infty$\\
$6(\alpha+1)$&$0.83$&$0.16$&$0.042$&$150.1$&$0.26$
\end{tabular}
\end{ruledtabular}
\end{table}

\begin{figure}[tbp]
\includegraphics[width=4.2cm]{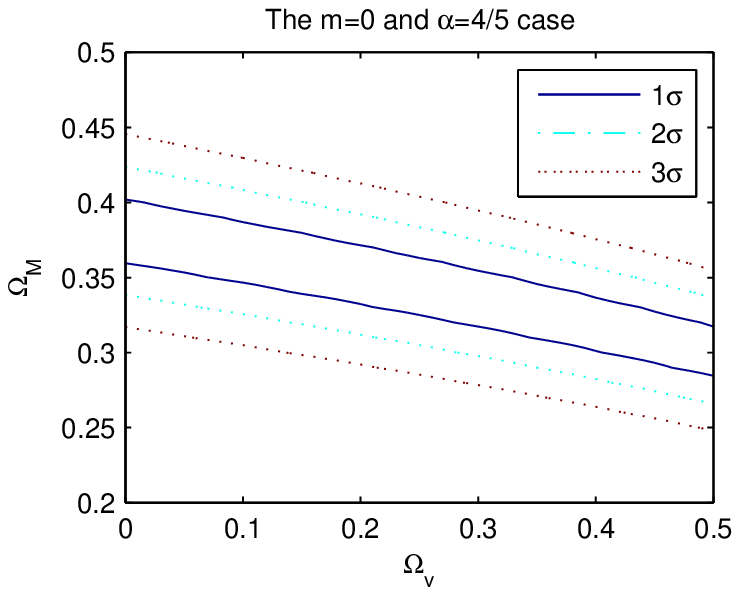} %
\includegraphics[width=4.2cm]{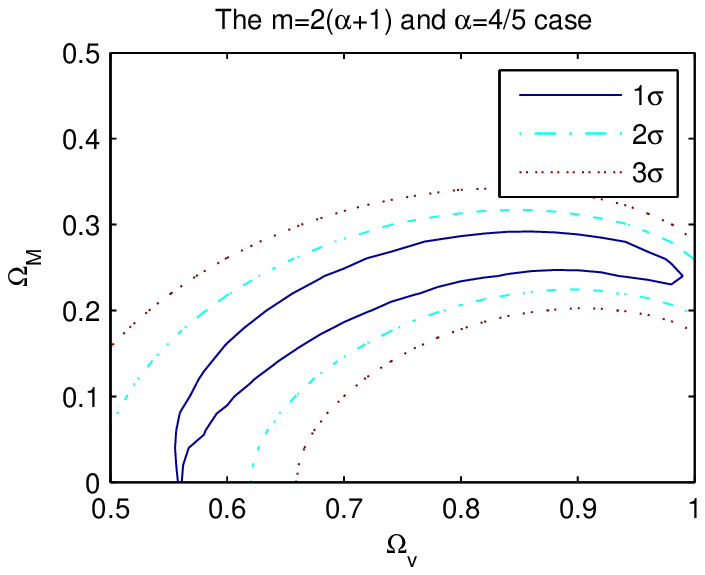} %
\includegraphics[width=4.2cm]{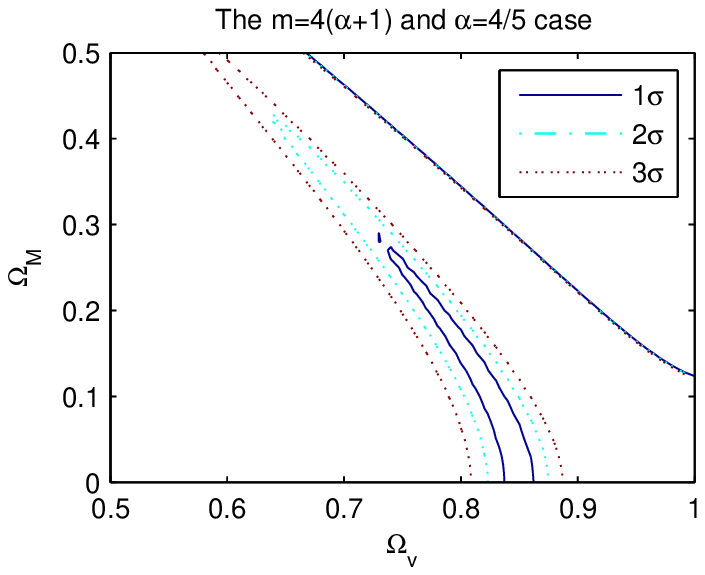} %
\includegraphics[width=4.2cm]{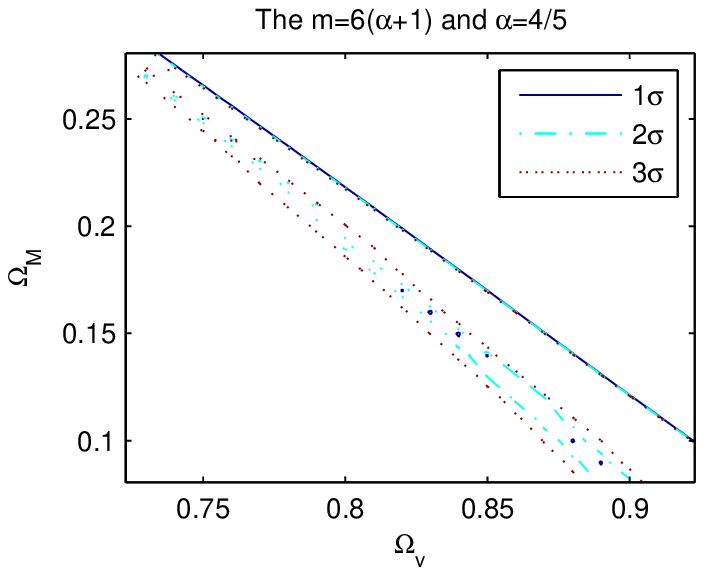}
\caption{The contours on the $\Omega_V-\Omega_M$ panel are in the case of $%
\protect\alpha=4/5$ for SNLS SNe and nearby data.}
\label{fig-SNLS3}
\end{figure}

\begin{table}[tbp]
\caption{The fitting results is in the case of $\protect\alpha=4/5$ for SNLS
SNe and nearby data. }
\label{tab4}%
\begin{ruledtabular}
\begin{tabular}{c|ccccc}
$m$&$\Omega_V$&$\Omega_M$&$\Omega_X$&$\chi^2$&$\eta$\\
\hline
$0$&$0.01$&$0.38$&$0.89$&$150.39$&$2.34$\\
$2(\alpha+1)$&$0.74$&$0.24$&$0.085$&$150.79$&$0.35$\\
$4(\alpha+1)$&$0.85$&$0$&$0.15$&$150.26$&$\infty$\\
$6(\alpha+1)$&$0.83$&$0.16$&$0.048$&$150.1$&$0.30$
\end{tabular}
\end{ruledtabular}
\end{table}

\begin{figure}[tbp]
\includegraphics[width=4.2cm]{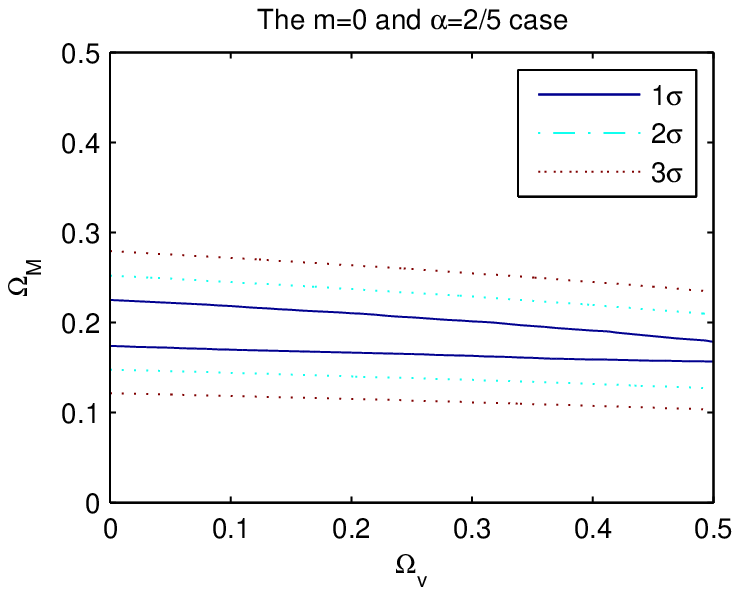} %
\includegraphics[width=4.2cm]{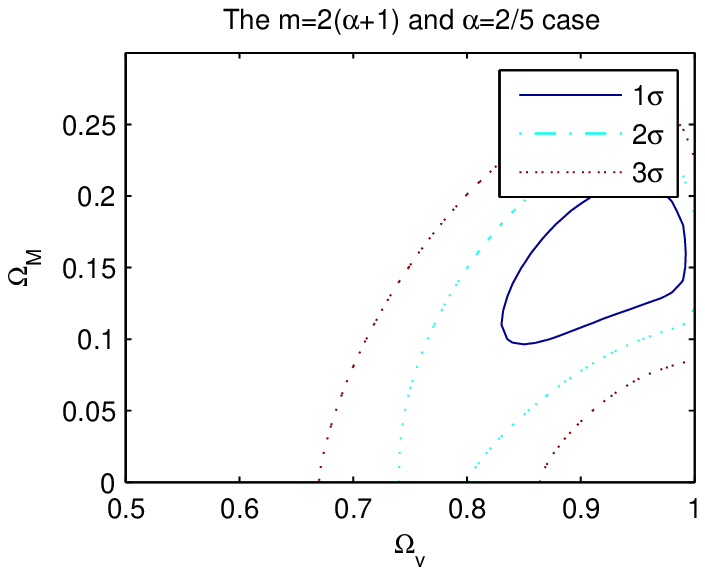} %
\includegraphics[width=4.2cm]{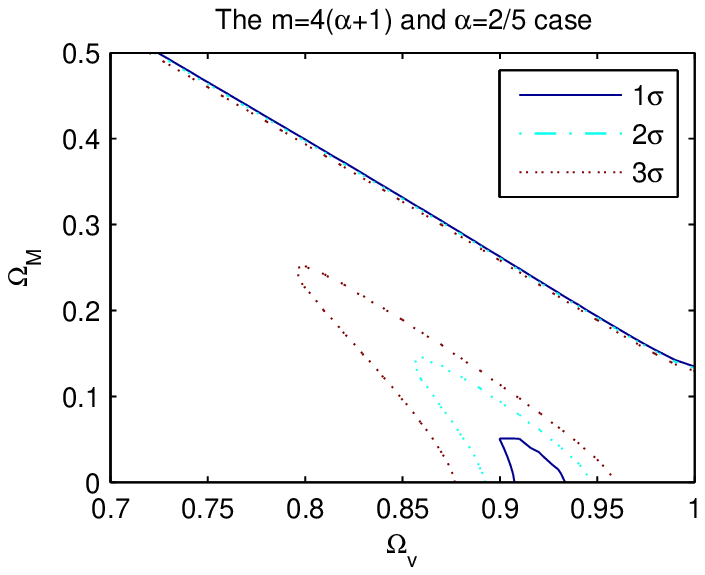} %
\includegraphics[width=4.2cm]{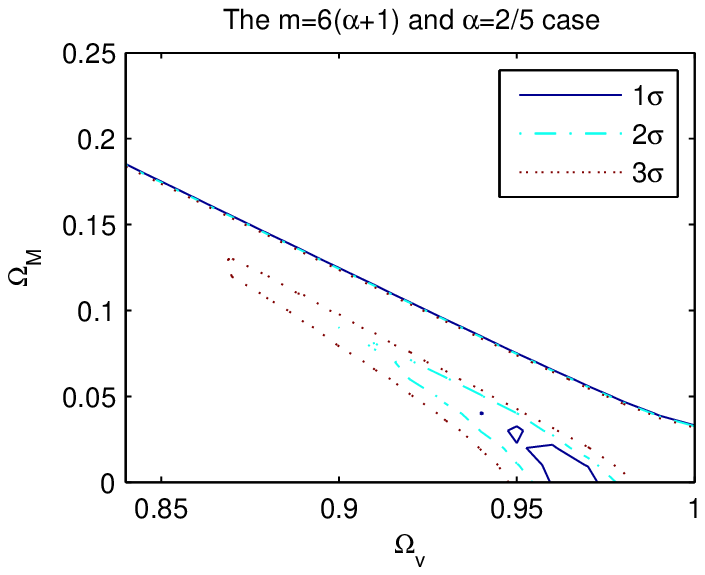}
\caption{The contours on the $\Omega_V-\Omega_M$ panel are in the case of $%
\protect\alpha=2/5$ for the 157 SN Ia gold data.)}
\label{fig-SNI1}
\end{figure}

\begin{table}[tbp]
\caption{The fitting results from the 157 gold data is in the case of $%
\protect\alpha=2/5$. }
\label{tab5}%
\begin{ruledtabular}
\begin{tabular}{c|ccccc}
$m$&$\Omega_V$&$\Omega_M$&$\Omega_X$&$\chi^2$&$\eta$\\
\hline
$0$&$0$&$0.2$&$0.92$&$213.97$&$4.62$\\
$2(\alpha+1)$&$0.87$&$0.13$&$0$&$216.95$&$0$\\
$4(\alpha+1)$&$0.92$&$0$&$0.08$&$213.55$&$\infty$\\
$6(\alpha+1)$&$0.96$&$0.02$&$0.029$&$208.2$&$1.42$
\end{tabular}
\end{ruledtabular}
\end{table}

\begin{figure}[tbp]
\includegraphics{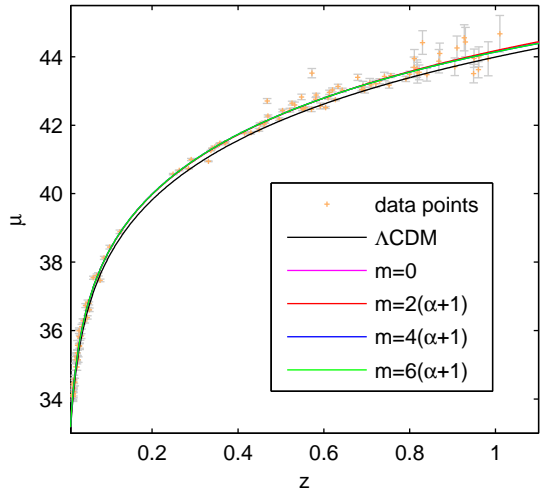}
\caption{Hubble diagram of SNLS and nearby SNe Ia}
\label{fig-SNLSbar}
\end{figure}

As a result, the fitting values of parameters are arranged at the Tabs. [\ref%
{tab2}, \ref{tab3}, \ref{tab4}, \ref{tab5}], the contours on the $%
\Omega_V-\Omega_M$ panel are displayed on Figs. [ \ref{fig-SNLS1}, \ref%
{fig-SNLS2}, \ref{fig-SNLS3}, \ref{fig-SNI1}] and at last the Hubble diagram
of SNLS SNe and nearby data is showed on the Fig. \ref{fig-SNLSbar}.

\subsection{analysis}

Results from the above tables have demonstrated that the best fitted values
for parameters except for $\eta $ are little affected by values of $\alpha $%
, but to the contrary, by values of $m$ and it shows obvious distinctions
between the different values of parameter $m$.
 We can rewrite the entangled degree into the form expressed
by relative ratio $\eta $ for convenience and brevity:
\begin{equation*}
P=\frac{2\eta ^{\alpha +1}}{\eta ^{2(\alpha +1)}+1},
\end{equation*}
which takes zero, with meaning no entanglement, at two limit cases
of either $\eta=0$ or $\eta=\infty$ obviously.

At the case $m=0$, the ECG model at low redshift from the Tab.
\ref{tab2} for instance has got the effective state parameter,
$w_{eff}\sim -0.8$, which is consistent with the usual range
$-1.3\lesssim w_{eff}\lesssim -0.8$, meanwhile it has returned to
the GCG model. In the Ref. \cite{Makler(2003)}
the GCG model has been considered by the use of Ia SNe with which
our result of $\Omega _{X}\sim 0.80$ is well consistent, that is,
$0.6\lesssim \Omega _{A}\lesssim 0.85$. $\Omega _{V}\sim 0$ may
have an implication that the dark side of universe favors an
entangled "cosmological constant" rather than the pure vacuum
energy. The entangled degree in this case is $P=0.23$ (case
$\alpha =2/5$ for example).

At the case $m=2(\alpha +1)$ which corresponds to curvature term,
$\Omega _{X}$ is a small but positive quantity. We can conclude
that in the ECG universe model by data fitting it tends to be
zero, that is a flat geometry and formally it is similar to the
well known $\Lambda$CDM model. The effective state parameter is
still $w_{eff}\sim -0.8$. The entangled degree is $P=0$ ($\alpha
=2/5$ case for example).

At the case $m=4(\alpha +1)$, that $\Omega _{M}=0$ corresponds to
the early stage of universe at which the radiation dominated. The
fitted parameters favor large value of dark energy or vacuum
energy and it has even existed since the earlier universe. Its
effective state parameter is $ w=-0.845$. The entangled degree is
$P=0$ ($\alpha =2/5$ for example) and there is no entanglement.

At the case $m=6(\alpha +1)$, the non-relativistic matter and
stiff matter can coexist with a relative larger vacuum energy
term. The entangled degree is $P=0.89$ ($\alpha =2/5$ for
example).

From the Hubble diagrams as showed on the Fig. \ref{fig-SNLSbar}, the ECG
models appear to be more consistent with data than the economic $\Lambda $%
CDM. It means that the entangled model of ECG is indeed an
excellent alternative to explain the currently cosmic expansion
speed-up. It is interesting to take the case of $m=0$ into
consideration, which depicts a
cosmological constant entangled with the matter phase similar to $\Lambda $%
CDM model, but may help to overcome, as suggested in the Ref.
\cite{chimento(2003)} the coincidence problem, which is the fatal
flaw in the$\Lambda $CDM model.

In this work with considerations of recent observational SNLS and
WMAP year three dataset analysis  we purposely investigate a more
practical model, an extended Chaplygin gas model or we may say a
$\Lambda$ plus entangled CDM model in which the entangled
X-component can be a radiation term, or curvature contribution, or
stiff matter or constant term in special cases. Actually in the
real cosmos those contributions may all exist but make different
effects by taking accordingly different fractions in universe
evolution stages.


\section{Conclusion and discussion}

\label{sec:Conclusion}

In this paper, in order to further discuss the properties of dark
energy which is used to explain currently cosmic accelerating
expansion, we have extended the \emph{Chaplygin gas} model by
replacing the parameter, $A$, with a possibly variable term.
Governed by the Friedmann equations, it shows a possibility to
describe an entangled state or unification between a matter phase
and any other phases like an entangled cosmological constant,
curvature, stiff matter term or the radiation contributions,
rather than only the vacuum energy component, which is the
character of the GCG model.

When the relation  $m=3(\alpha +1)$ holds, one finds that the ECG
model can be treated as the $\Lambda $CDM at the low redshift
cases. On another side, as the inequation, $m\neq 3(\alpha +1)$,
exists, the ECG model realizes what we have expected about the
extension to GCG model. Further, the likelihood function is used
to check whether such modification is reasonable. With the fitted
results and analysis on the above sections, we can not exclude the
possibility that there are indeed some other phases which can
successfully entangle with the matter phase to determine the fate
of our universe evolution altogether.

The fact that the GCGM which is deduced for the phenomenological
reasons can be interpreted from a brane-world view or the
Born-Infeld theory as shown below.
The field $\phi $ and the energy density are related by the expression
\begin{equation}
\phi ^{2}(\rho )=\rho ^{\alpha }(\rho ^{1+\alpha }-A)^{(1-\alpha )/(1+\alpha
)},  \label{1:field}
\end{equation}%
with the Lagrangian density as
\begin{equation}
\mathcal{L}_{GBI}=-A^{1/(1+\alpha )}\left[ 1-(g^{\mu \nu }\theta _{,\mu
}\theta _{,\nu })^{(1+\alpha )/2\alpha }\right] ^{\alpha /(1+\alpha )},
\label{1:lagran}
\end{equation}%
(more details can be found in the Ref. \cite{Bento:2002}). Likewise, it is
possible to construct such relation corresponding to the ECG model. The
Lagrangian density for a massive complex scalar field, $\Phi $, is
\begin{equation*}
\mathcal{L}=g^{\mu \nu }\Phi _{,\mu }^{\ast }\Phi _{,\nu }-V(|\Phi |^{2}),
\end{equation*}%
which is suggestive in the Ref. \cite{Bilic(2002)}, and the field can be
expressed in terms of its mass, $m$, as $\Phi \equiv (\phi /\sqrt{2}m)\exp
(-im\theta )$. Adopting the method in the Ref. \cite{Bento:2002}, we here
display the Lagrangian density for the ECG model.


The density and pressure have had the relations as
\begin{equation}
\rho =\frac{\phi ^{2}}{2}V^{\prime }+V,\quad p=\frac{\phi ^{2}}{2}V^{\prime
}-V,  \label{2.2:rho-field}
\end{equation}%
where $V=V(\phi /2)$ and $V^{\prime }(x)=dV/dx$. The resulting
Lagrangian density for GCGM which may have got a brane connection
is
\begin{equation}
\mathcal{L}_{GBI}=-A^{1/(1+\alpha )}\left[ 1-(g^{\mu \nu }\theta _{,\mu
}\theta _{,\nu })^{(1+\alpha )/2\alpha }\right] ^{\alpha /(1+\alpha )}.
\label{2.2:lagran}
\end{equation}%
At the cases of low redshifts, it is admissible that we just
replace the constant parameter, $A$, in the Eq. (\ref{2.2:lagran})
by the possibly variable expression, $A_{0}(1+z)^{m}$, to obtain
the Lagrangian density for the ECG
model. In another way, from the Eqs. [ (\ref{2:EOS}) and (\ref{2.2:rho-field}%
) ] we can derive
\begin{equation}
(1+z)^{m}=-\frac{1}{A_{0}}\left( \frac{\phi ^{2}}{2}V^{\prime }-V\right)
\left( \frac{\phi ^{2}}{2}V^{\prime }+V\right) ^{\alpha }.
\label{2.2:z-field}
\end{equation}%
Evidently, it suffices to write the Eq. (\ref{2.2:lagran}) into the form
\begin{equation}
\mathcal{L}_{GBI}=-V_{ECG}\left[ 1-(g^{\mu \nu }\theta _{,\mu }\theta _{,\nu
})^{(1+\alpha )/2\alpha }\right] ^{\alpha /(1+\alpha )},  \label{2.2:lagran2}
\end{equation}%
where $V_{ECG}$ is an effective potential function. When the $\alpha =1$ it
still can reproduce the Born-Infeld Lagrangian density 

Observational cosmology has challenged our naive physics models,
and with the anticipated advent of more precious data we will have
the chance to understand or uncover the universe mysteries by more
practical modelling. Quite possibly we will get more hints to
unveil the cloudy cosmological constant puzzle and test whether
there are really the mixed states or unified dark energy and
matter or dark fluid in reasonable
 universe evolution.
\begin{acknowledgements}
We thank Prof. S.D. Odintsov  for reading the manuscript with
helpful comments and Profs. I. Brevik and Lewis H.Ryder for lots of
interesting discussions. This work is partly supported by NSF and
Doctoral Foundation of China.
\end{acknowledgements}

\end{document}